\documentclass[
  11pt,
  prd,
   onecolumn,
  superscriptaddress,
  nofootinbib,
  amsmath,
  amssymb,
  aps,
]{revtex4-2}

\pdfoutput=1

\usepackage[utf8]{inputenc}
\usepackage{graphicx}
\usepackage{bm}
\usepackage{microtype}
\usepackage{threeparttable}
\usepackage{mathtools}
\usepackage[colorlinks=true]{hyperref}
\usepackage[normalem]{ulem}

\usepackage{cleveref}
\crefname{figure}{Fig.}{Figure}
\crefname{equation}{Eq.}{Equation}
\crefname{equations}{Eqs.}{Equations}
\crefname{appendix}{App.}{Appendix}

\newcommand{\f}[2]{\frac{#1}{#2}}

\newcommand{\half}{{\frac{1}{2}}}

\newcommand{\symmm}{${\mathcal N}=4$ supersymmetric Yang-Mills theory}

\newcommand{\be}{
\begin{equation}}
  \newcommand{\ee}{
\end{equation}}

\newcommand{\ed}{{\cal E}}      
\newcommand{\pL}{{\cal P}_L}
\newcommand{\pT}{{\cal P}_T}
\newcommand{\pp}{{\cal P}}
\newcommand{\pa}{{\cal A}}

\newcommand{\mn}{{\mu\nu}}

\begin{document}

\author{Alexander Soloviev}
\email{alexander.soloviev@fmf.uni-lj.si}
\affiliation{Faculty of Mathematics and Physics, University of Ljubljana, Jadranska ulica 19, SI-1000, Ljubljana, Slovenia}

\author{Michał Spaliński}

\email{michal.spalinski@ncbj.gov.pl}
\affiliation{National Centre for Nuclear Research, 02-093 Warsaw, Poland}
\affiliation{Physics Department, University of Białystok,
15-245 Bia\l ystok, Poland}

\title{ Asymptotics of superfluid Bjorken flow }

\begin{abstract}
  We consider the dynamics of an expanding superfluid modeled by Mueller-Israel-Stewart theory coupled to a complex scalar field with a $U(1)$ symmetry that is spontaneously broken. 
  This is a manageable theoretical setting for explorations of the chiral phase transition of expanding quark-gluon plasma. We study the late proper-time behavior of Bjorken flow in this physical system and find that asymptotic solutions can be expressed as a transseries of a novel form, which contains factors like $\tau^{-a\ln \tau}$.  This transseries describes how the information encoded in the initial data is diluted in the course of dissipative evolution. These solutions retain memory of the symmetry-breaking transition and describe two qualitatively different late-time behaviors of the dynamical variables, depending on 
  condensate relaxation rate: 
 either a purely damped fall-off or damped oscillations. 
  The possibility that such oscillations could be imprinted in the observed outcomes of heavy ion collision experiments is the main physical insight that follows from our analysis. %
\end{abstract}

\maketitle

\section{Introduction}
\label{sec:intro}

Fluid dynamics can be viewed as a systematic effective field theory approach
using fundamental conservation laws to describe long wavelength excitations of
systems out of equilibrium \cite{Kovtun:2012rj,romatschke2019relativistic}. In certain settings, such as near critical loci in the phase diagram, these excitations include Goldstone modes of spontaneously broken symmetries, as occurs in
superfluids, including those present in nuclear matter \cite{Son:1999pa,Son:2002zn,Son:2002ci}.

The focus of this paper is on the physics of quark-gluon plasma (QGP) in the context of heavy ion collision experiments, where critical dynamics constitute an important part in modern experimental searches for the QCD critical point \cite{Busza:2018rrf}.
In this context it is useful to consider the idealized case of QCD in
the chiral limit, where spontaneous breaking of chiral symmetry leads to the
appearance of massless Goldstone bosons which can be identified with the pions.  Even though in the real world chiral symmetry is explicitly broken by quark masses, there are many features of QCD which can be understood from the perspective which views the explicit breaking of chiral symmetry in the light quark sector as a perturbation. This perspective has also been explored in the context of QGP physics, in part to explain the enhanced production of soft pions as a signature of the chiral phase transition~\cite{Grossi:2020ezz,Grossi:2021gqi,Florio:2021jlx,Florio:2023kmy,Florio:2025lvu,Florio:2025zqv}.

In this study, instead of considering the nonabelian chiral symmetry group of massless QCD,
we consider a simple model of a
complex scalar field with dynamics invariant under a $U(1)$ symmetry,
spontaneously broken by the ground state. The resulting Goldstone boson is
coupled to an expanding conformal fluid described using Mueller-Israel-Stewart (MIS) theory~\cite{mueller,israel}. Following earlier
work~\cite{Mitra:2020hbj,Buza:2024jxe}, we consider the combined system undergoing Bjorken flow, where the dynamics is captured by a system of $3$ ordinary differential equations. 
The non-dissipative version of this  theory has been shown to lead to causal and stable evolution in flat space, see \cite{Gavassino:2025zmy} for details.
The physical picture of the typical evolution of the Bjorken superfluid is as follows. In analogy to the QGP, the system is initialized with a temperature in the unbroken phase, i.e.~$T>T_c$. The condensate rapidly rolls down to its minimum value, which in the unbroken phase is $\rho=0.$ The temperature evolution in this regime is thus unaffected by the condensate, evolving in accordance with the MIS attractor\footnote{
In Ref.~\cite{Buza:2024jxe}, the term {\em attractor time} was coined to refer to the length of time that the system remains in this regime.
}.
Once the temperature drops below the critical value $T_c$, the mass term in the potential becomes negative, triggering spontaneous symmetry breaking and the condensate rolls down to the minimum of the potential in the broken phase. 
Despite the continued expansion, the system then freezes to a constant non-zero value the condensate, as well as a constant nonzero value of the temperature, in contrast to the typical Bjorken evolution
$T\propto\tau^{-1/3}$ at late times, see e.g. Ref.~\cite{Bjorken:1982qr}.

While earlier studies relied mostly on numerical work, here we apply asymptotic
methods to study the physics at late times.
The power series asymptotic behavior of the superfluid system studied here has novel features, notably the occurrence of logarithms. Their appearance can in fact be traced to a simplified system where the condensate is frozen to its asymptotic value; in that case one can reduce the equations of motion to a single second order ODE for the effective temperature, where logarithms appear in the asymptotics even in the case of an ideal fluid. 
The asymptotic late time solution 
can be seen to take the  form 
 \begin{equation}
 \label{eq:toyasym}
      T(\tau) \sim \sum_{n=0}^\infty \sum_{m=0}^n t_{n,m}\tau^{-n} \ln^m (\Lambda \tau),
 \end{equation}
where the $t_{n,m}$ are constant coefficients determined by the evolution equations and
$\Lambda$ is an integration constant. It turns out that the asymptotic solution of the full superfluid system retains the form of \cref{eq:toyasym}.
While the form of this series may appear unusual,
series of this form have appeared in other contexts
-- further details are provided in \cref{app:logs}.

It is well known that the late proper time expansion is divergent and captures
the physics in a rather subtle way, with a part of the information carried by
power law terms, while the rest is exponentially suppressed~\cite{Heller:2015dha} (see also the reviews~\cite{Florkowski:2017olj,Soloviev:2021lhs,Jankowski:2023fdz}).
These exponentially suppressed terms are in correspondence with the quasinormal
modes of the microscopic system.
In the case of systems previously studied, these modes were connected with relaxation of the shear stress and were necessary to ensure the causality as well as the stability of the equilibrium state~\cite{Spalinski:2016fnj,Heller:2022ejw}. In the superfluid system the nonhydrodynamic quasinormal modes include ones connected to the relaxation of the symmetry breaking condensate, and depend on its relaxation rate, denoted by $C_\kappa$ below. 
The mathematical
structure of late time asymptotic solutions is that of a transseries, and in this work we establish its form in the case of superfluid Bjorken flow. This
transseries turns out to have novel features, but expresses the same physical idea as in other examples~\cite{Heller:2015dha,Aniceto:2015mto,Aniceto:2018uik,Aniceto:2022dnm}.
A physically interesting point which emerges from the present analysis is that 
the exponentially damped terms in the superfluid system can include oscillating
modes which describe the dynamics of the condensate. The precise form of the
transseries reveals that there is a critical value of the condensate relaxation rate below which oscillations are present. Outside this parameter domain the
system becomes overdamped and the oscillations are no longer present. This fact can be verified numerically and such analysis confirms the conclusions implied by the transseries solution.
If present, such oscillations could leave an imprint in the observables measured in heavy-ion collision experiments.

The outline of the paper is as follows: in \Cref{sec:setup}, we outline the superfluid construction in Bjorken flow. We then discuss the frozen condensate limit in \Cref{sec:frozen}, where we are firmly in the broken regime, but the condensate dynamics play no role. This serves as a warm up to the full dynamical case in \Cref{sec:unfrozen}, where we analyze the asymptotics of the resulting theory. In \Cref{sec:large}, we present some results concerning the large order behavior of the series appearing in the lowest transseries sectors and show that the pattern of  branch point singularities in the Borel plane matches the structure of the transseries.

\section{Superfluid Bjorken flow}\label{sec:setup}

The chiral symmetry group of QCD with two massless flavors of quarks is  $SU(2)\times SU(2)\sim O(4)$. Spontaneous breaking of this symmetry, signaled by a nonzero value of the quark condensate, leads to the appearance of a triplet of Goldstone bosons that are interpreted as the chiral limit analog of the physical pion triplet. In this paper, following~\cite{Mitra:2020hbj}, we consider a simpler model involving a global $U(1)$ symmetry, where only one Goldstone boson appears. Obviously,  this cannot be used directly as a model of pion physics, but it may serve as an illustration for the dynamics of the condensate interacting with the QGP, which is the focus of interest here.

In the ideal fluid limit, the system of interest can be described by the effective action \cite{Mitra:2020hbj,Buza:2024jxe}
\begin{equation}\label{eq:action}
S=\int d^4 x \sqrt{-g}\left[ p(T) -\frac{1}{2} (D_\mu \Sigma)(D^\mu \Sigma)^\dagger - V( \Sigma, T, \mu)\right]\,,
\end{equation}
where $\Sigma= \rho e^{i\psi}$ is the complex scalar field with $\rho$ denoting the condensate and $\psi$ the $U(1)$ phase. The potential is given by 
\begin{align}\label{potential}
V( \rho,  T, \mu) = \frac{m_0 (T-T_c)}{2}  \rho^2 + \frac{\lambda}{4}  \rho^4.
\end{align}
Here $T$ is the (dynamical) effective temperature and $T_c$ is the critical temperature. Note that the mass term explicitly changes sign as the system temperature drops below the critical temperature and into the broken phase.
In what follows, we will concern ourselves with the dynamics of the condensate, $\rho,$ at zero chemical potential, which due to the Josephson constraint fixes the phase to be a constant (see \cite{Mitra:2020hbj} for the formulation of the present theory with a dynamical phase).

The equations of motion express the conservation of energy-momentum, the dynamics of the order parameter and the relaxation of the shear-stress according to MIS theory. 
The energy momentum tensor,  
$T^\mn= T^\mn_{\rm ideal}+ \Pi^\mn,$ decomposed into ideal and dissipative contributions respectively, is conserved
\begin{align}
\nabla_\mu T^\mn&=0,\label{eq:emt}\\
T^{\mu\nu}_{\rm ideal}&\equiv\frac{2}{\sqrt{-g}}\frac{\delta S}{\delta g_{\mu\nu}} = 
\varepsilon \, u^\mu u^\nu + p \, \Delta^{\mu\nu}
+\nabla^\mu  \rho \nabla^\nu  \rho 
-g^\mn \left(\frac{1}{2}\nabla_\mu \rho \nabla^\mu \rho + V\right),
\end{align}
where 
$\varepsilon$ is the energy density, $p$ is the pressure and
$u^\mu$ is the four velocity (timelike normalized $u^\mu u_\mu=-1)$ and  $\Delta^\mn=u^\mu u^\nu+g^\mn.$ Note that the dissipative tensor is both orthogonal to the fluid velocity and traceless, i.e.~$\Pi^\mn u_\mu=0=\Pi^\mn g_\mn.$
In what follows, we will consider a conformal fluid with the equation of state, $\varepsilon=3p\sim T^4$.

The equation of motion for the order parameter is assumed to be
\begin{align}\label{eq:cond}
 -\nabla_\mu \nabla^\mu  \rho  +\frac{\partial V}{\partial  \rho}
 &= 
 -\kappa u^\mu\nabla_\mu  \rho.
\end{align}
This differs form the Euler-Lagrange equation following from the action \Cref{eq:action} by the addition of the dissipative term on the right hand side. This term describes the relaxation of the condensate, 
where $\kappa$ is a transport coefficient which captures the relaxation rate of the condensate to the minimum of the potential.
It will be shown below that this terms affects the asymptotic late time behavior by damping oscillations of the condensate.

The final element is the MIS relaxation equation \cite{mueller,israel,Baier:2007ix} 
\begin{equation}
    \tau_\pi \left(u\cdot\nabla + \frac{4}{3} \, \nabla \cdot u\right) \Pi^{\mu\nu}+ \Pi^{\mu\nu} = -  2 \, \eta \, \sigma^{\mu\nu},
    \label{eq:mis}
\end{equation}
where $\eta$ is the shear viscosity, $\tau_\pi$ is the relaxation timescale and 
\begin{equation}
    \sigma^\mn =\frac{1}{2} \Delta^{\mu\alpha}\Delta^{\nu\beta}(\nabla_\alpha u_\beta+\nabla_\beta u_\alpha)-\frac{1}{3}\Delta^\mn\nabla_\alpha u^\alpha
\end{equation}
is the shear tensor.
We adopt the temperature dependence of the transport coefficients appropriate for high temperatures
\begin{align}
\eta = \frac{4}{3}C_\eta T^{3},  \quad \,\,\, \tau_\pi = C_{\tau} T^{-1},\quad 
\kappa = C_{\kappa} T.
\end{align} At lower temperatures, one would expect these to be modified by mass corrections, $m_0$,
see footnote 2 in \cite{Mitra:2020hbj}. 

In our study we consider the system undergoing Bjorken flow, which captures a number of kinematic as well as dynamical features of the  QGP created in heavy ion collisions. 
The symmetries of Bjorken flow~\cite{Bjorken:1982qr} are implemented in the standard way by passing to Milne coordinates $(\tau, \eta)$ 
with the metric 
\begin{align}
    ds^2=-d\tau^2+dx_\perp^2+\tau^2 d\eta^2,
\end{align} 
and imposing the condition that all fields depend only on $\tau$.
We parameterize the dissipation tensor 
via the dimensionless quantity 
$\chi$ defined by $\Pi^{\eta}_\eta = -\chi(\varepsilon+p)$. 
We can relate this to pressure anisotropy of Bjorken flow,  defined in terms of eigenvalues of the
  energy-momentum tensor $T_{\phantom{\mu}\nu}^\mu \equiv \mathrm{diag} ( -\ed, \pT, \pT,\pL)$ using 
\begin{equation}
       \pa \equiv \frac{\pT - \pL}{\pp},
       \qquad
       \pp = \f{1}{3}(\pL + 2\\\pT),
  \end{equation}  
can be expressed as $\pa =\frac{3}{2}\frac{\varepsilon+p}{\pp}\chi $.

The dissipative equations of motion \eqref{eq:cond}, \eqref{eq:emt} and \eqref{eq:mis} then take the form \cite{Mitra:2020hbj,Buza:2024jxe}
\begin{subequations}
\label{eq:eoms}
\begin{align}
\rho'' &+ \frac{\rho'}{\tau} + \lambda \rho^3 + m_0(T- T_c) \rho 
= - {C_{\kappa}}T\rho', \\
\frac{\tau T'}{T}  &=\frac{1}{3} (\chi- 1)+m_0\frac{\rho^2 + 2 \tau \rho \rho'}{8T^3}
+ \frac{\tau}{4T^3}
C_{\kappa}{\rho'}^2
,\\
\label{Eq:ChiEq}
\tau \chi' &+ \frac{4}{3} \left(\chi - \frac{C_\eta}{C_\tau} \right)+4 \chi \frac{\tau T'}{T} + \frac{\tau}{C_\tau} \chi T = 0,
\end{align}
\end{subequations}
where the prime denotes the derivative with respect to $\tau,$ e.g.~$\rho^\prime = \partial_\tau \rho.$

In the above discussion, we have neglected the dynamics of the $U(1)$ phase. Note that in the current homogeneous Bjorken setting, it was shown in \cite{Mitra:2020hbj} that the phase dynamics decouple early 
due to the expansion. 
Since at present we are primarily concerned with homogeneous late time dynamics, we will work at zero chemical potential.
Note also that
the Bjorken expansion acts in some sense as a non-linear quench, so in the expanding $O(4)$ theory we expect that the pion dynamics will play an important role at later times, as in  simulations in the quenched scenarios \cite{Florio:2025lvu,Florio:2025zqv}.

In order to simplify analytic calculations, we will adopt a specific choice of potential parameters. 
A convenient choice which we will use unless noted otherwise is given by
\begin{align}\label{eq:nice-parameters}
\lambda=\frac{1}{6 T_c} ,  \qquad  
m_0=\frac{T_c}{3}.
\end{align}
In what follows, we also 
set $T_c=1$, i.e., we measure in units of $T_c$. These choices are motivated only by the desire to 
display 
formulae in a readable form. The results presented in the following
can also be obtained for general values of the parameters appearing in the potential; some results that do not assume the values in \cref{eq:nice-parameters} are presented in \cref{app:general}.

For analytic computations, we 
will keep the 
transport coefficients arbitrary. However, in numerical calculations we will take 
\begin{align}
    C_\eta = \frac{1}{4\pi}, 
    \qquad
    C_\tau = \frac{2-\ln 2}{2\pi},
\end{align}
which are the values appropriate for \symmm~\cite{Bhattacharyya:2007vjd}. 
We will consider various values 
of
 $C_\kappa$, since the superfluid system exhibits qualitatively different behavior depending on the value of this parameter, as we show in the following.

\section{The frozen condensate limit}\label{sec:frozen}

Before 
undertaking the analysis of the complete set of equations \eqref{eq:eoms}, it will be helpful to first gain some intuition by considering the 
limit
$C_\kappa \rightarrow \infty$, in which the condensate assumes a constant value $\rho(\tau) = \rho_0$ 
and the evolution equations simplify considerably. To get a consistent set of equations one also needs to take a suitable limit for parameters appearing in the potential energy, such that $m_0\to0$ with $m_0 \rho_0^2$ remaining constant. 
This reduces the order of the equations of motion for $T=T(\tau)$ and leads to 
\begin{align}
\label{eq:toy}
    \frac{C_\tau  \tau  T''}{T}+\frac{3
   C_\tau  \tau  (T')^2}{T^2} &+\left(\frac{11 C_\tau }{3 T}+\tau \right)
   T'+\frac{T}{3}  
   -\frac{4 (C_\eta -C_\tau )}{9 \tau }
= \xi^3  \frac{\left(3 C_\tau  \tau  T'+4 C_\tau  T+3 \tau
   T^2\right)}{72 \tau  T^4},
\end{align}
where $\xi^3\equiv3 \rho^2_0 m_0$.
If the constant value of the condensate $\rho_0$ is set to zero, this is 
the same equation for the effective temperature as considered in Ref.~\cite{Heller:2015dha}.

To build intuition, we first consider the even simpler Navier-Stokes limit, $C_\tau\rightarrow 0$, of \cref{eq:toy}. In the zero condensate case, $\xi\rightarrow0,$ the solution is the well-known 
exact solution for viscous Bjorken flow 
\begin{align}
   T(\tau)= \frac{\Lambda}{(\Lambda\tau)^{1/3}}
    \left( 1 -\frac{2
   C_\eta }{3}\frac{1}{(\Lambda\tau)^{2/3}}
   \right),
\end{align}
where $\Lambda$ is a constant of integration. 
For a non-vanishing condensate, the Navier-Stokes equation for Bjorken flow cannot be solved in closed form. One can however calculate the late time asymptotic solution, which takes the form of \cref{eq:toyasym}; specifically
\begin{align}
\label{eq:NS}
    T(\tau)=\frac{\xi }{2}
    +\frac{4
   C_\eta \ln (\Lambda\tau)}{9 \tau }
   -\left(2 +2
   \ln (\Lambda\tau) +  \ln^2(\Lambda\tau)\right)\frac{32 C_\eta^2}{81\xi}\tau^{-2}
   +\mathcal{O}(\tau^{-3}),
\end{align}
where $\Lambda$ is again an integration constant. 
We see that 
now the temperature does not vanish at late times, with the limiting value set by the condensate.
Furthermore, the late time expansion contains all negative integer powers of the proper time, but the result is no longer just a power series: we see a novel structure where the coefficients of powers of $\tau$ are  polynomials in $\ln(\Lambda \tau)$.

We now return to \cref{eq:toy} with non-zero $C_\tau$. 
The asymptotic late time solution can again  be seen to take the form  given in \cref{eq:toyasym}.
The coefficients appearing in \cref{eq:toyasym} are determined by \cref{eq:toy}, apart from the constant $t_{1,0}$ which remains undetermined and is thus a constant of  integration. 
One can however absorb it into the scale of the logarithm, 
so that $\Lambda$ is in fact the only integration constant that appears in the asymptotic solution.
At leading orders one finds
\begin{align}
    t_{0,0} = \half,\ t_{1,1} = \f{4 C_\eta}{9}, \ t_{2,2} = -\frac{32 C_\eta^2}{81},\ t_{2,1} =-\frac{64 C_\eta^2}{81} ,\ t_{2,0} = -\frac{8 }{81}\left(8 C_\eta^2-3 C_\eta C_\tau\right).
\end{align}

\begin{figure}
    \centering
\includegraphics[width=0.7\linewidth]{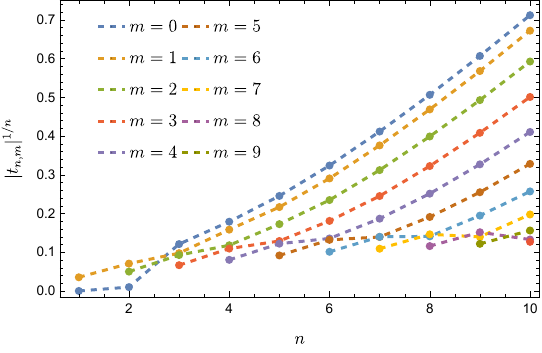}
    \caption{
    The factorial growth of the coefficients in \cref{eq:toyasym} of the frozen condensate case, where each curve corresponds to a specific log-sector, with  increasing power of $\ln (\Lambda\tau)$ going from left to right.
    }
    \label{fig:simple-n}
\end{figure}

 The asymptotic series 
 \cref{eq:toyasym}
 can be rearranged into sectors grouping terms proportional to a given power of the logarithm: 
\begin{equation}
 \label{eq:toysectors}
      T(\tau) \sim \sum_{m=0}^\infty \ln^m (\Lambda \tau) \left[\sum_{n=0}^\infty t_{n,m}\tau^{-n} \right],
 \end{equation}
 where we have defined $t_{n,m}=0$ for $m>n$. 
 The power series appearing in each log-sector can be seen to have a vanishing radius of convergence; they are in fact factorially divergent, as seen in  \cref{fig:simple-n}. This is typical of late proper time expansions in Bjorken flow, and the hydrodynamic gradient expansion in general.

As is well known, asymptotic series 
can
be considered as elements of a transseries which include terms beyond all orders of the original expansion. In the context of relativistic  hydrodynamics such contributions are exponentially suppressed and depend on the nonhydrodynamic modes which are required by causality. In the present case they also depend on the symmetry breaking sector. 
Proceeding as in Ref.~\cite{Heller:2015dha}, 
one finds that the first transseries sector assumes the form
\begin{align}
\label{eq:toyts}
    \delta T(\tau) \sim \sigma {\tau^{-7/3} e^{-\frac{\xi \tau }{2 C_\tau }-\frac{2 C_\eta  \ln^2(\Lambda \tau )}{9 C_\tau }}
   \left(\xi - \frac{32 C_\eta ^2 }{81 C_\tau  } \frac{\ln^2(\Lambda \tau )}{ \tau} - 
   \frac{32 C_\eta  \left(4 C_\eta +9  C_\tau \right) }{81 C_\tau} 
   \frac{\ln(\Lambda \tau )}{ \tau}
   \dots\right)}
\end{align}
The constant $\sigma$ is the transseries parameter which plays the role of the second  integration constant of \cref{eq:toy}. 
The 
 series
 appearing in \cref{eq:toysectors} as well as the transseries sector given in \cref{eq:toyts} have a novel form compared to other cases studied in the context of Bjorken flow.
The appearance of logarithms in the perturbative sector of a transseries was previously noted for instance in the case of the Painlev\'e I equation~\cite{Aniceto:2011nu}, but to the best of our knowledge the $\ln^2(\Lambda\tau)$ term which appears in the transseries element in \cref{eq:toyts} has not been observed previously.  
The expansion coefficients appearing in the infinite series above are expected to satisfy resurgence relations which connect them to the late coefficients of the perturbative series  
\cref{eq:toysectors},
in a similar way to what happens in the case without a condensate. These relations are essential for an ambiguity-free result of Borel resummation applied sector by sector to the full transseries. 
We leave a careful analysis of these matters    to future work.

\section{Late time asymptotics}\label{sec:unfrozen}

Having determined the form of the late time dynamics in the simpler setting of the frozen condensate limit in the previous section, we now turn our attention to the full Bjorken superfluid asymptotics. Numerical solutions to \cref{eq:eoms} have been previously studied in Refs.~\cite{Mitra:2020hbj,Buza:2024jxe}. Here we turn to the main focus of the present article, which is an analytic description of these solutions at late time. 

The form of the
asymptotic solutions is found to be 
\begin{align}
\label{eq:asymptotic}
   T(\tau) &\sim \sum_{n=0}^\infty \sum_{m=0}^n t_{n,m}\tau^{-n} \ln^m (\Lambda \tau) \equiv \Phi^{(0)}(\tau),
\end{align}
with asymptotic solutions for $\rho$ and $\chi$ of the same structure, 
with coefficients $r_{mn}$ and $x_{mn}$. 
These coefficients 
can be determined from the equations of motion.
Adopting the parameter values listed in \cref{eq:nice-parameters} leads to the leading order values of $t_{0,0}=1/2$, $x_{0,0}=0$ and $r_{0,0}=1.$ Full expressions with arbitrary parameters can be found in \Cref{app:general}.
Note that 
$t_{1,0}$
is an arbitrary integration constant\footnote{Note that we could instead choose e.g.~$r_{1,0}$ as the integration constant.
}; 
as in the case of the frozen condensate in \Cref{sec:frozen}, 
it can absorbed into the logarithms
--- this has already been done in \cref{eq:asymptotic}, where this integration constant is denoted by $\Lambda$. 
We will return to the infinite series $\Phi^{(0)}$ in the following section.

The constant $\Lambda$ is the only part of the initial data of the system \cref{eq:eoms} that survives to asymptotically late times. 
The remaining initial data to which solutions of \cref{eq:eoms} are entitled appears in the form of contributions which are beyond the gradient expansion, that is, contributions which are suppressed more strongly than any power of the proper time. Such contributions enter as exponentially damped corrections to classical asymptotics, with 
the complex frequencies that 
 are in one to one correspondence with nonhydrodynamic quasinormal modes of the system.
In the case at hand
one finds
\begin{align}
\label{eq:transseries-temp}
T(\tau) &\sim \Phi^{(0)}(\tau) + \sum_{k=0, \pm}\sigma_k\tau^{\beta_k}\,
e^{
-\gamma_k \tau - \gamma'_k\ln^2(\Lambda\tau )}  \Phi^{(1)}_k(\tau) + \ldots
\end{align}
The series 
\begin{align}
   \Phi^{(1)}_k \sim 1 + \mathcal{O}\left(\frac{\ln(\Lambda\tau)^2}{\tau}\right)
\end{align}
are infinite series defining the first transseries sector. There are of course
further such contributions, as is always the case for nonlinear systems. One can obtain the transseries sectors for $\rho$ and $\chi$ in the same way as one can obtain the perturbative sectors for them from the perturbative series $\Phi^{(0)}$.

The parameters of the purely damped sector are given by
\begin{align}
\label{eq:ts0}
\beta_0 = -\frac{7}{3},\quad
\gamma_0 = \frac{1 }{2 C_\tau } \quad
\gamma'_0 = \frac{C_\eta }{6 C_\tau } ,
\end{align}
and parameters of the oscillatory sectors are 
\begin{align}
  \beta_\pm &= - \frac{1}{2} \pm \frac{3 C_\kappa }{2 \sqrt{9 C_\kappa ^2-64}}, \quad
   \gamma_\pm= \frac{C_\kappa }{4} \pm \frac{1}{12} \sqrt{9 C_\kappa ^2-64},
   \quad
   \gamma'_\pm =
   \frac{C_\eta }{12} \left( C_\kappa 
 \pm  \frac{ 3 C_\kappa ^2+16}{\sqrt{9 C_\kappa
   ^2-64}}\right).
   \label{eq:tspm}
\end{align}
Equations \eqref{eq:ts0} and \eqref{eq:tspm} describe the leading contributions to a transseries solution of \cref{eq:eoms}, with the power law asymptotic series $\Phi^{(0)}(\tau)$ acting as the perturbative sector.
The coefficients $\gamma_k$ appearing in the exponentials coincide with the location of branch point singularities in the Borel plane, as seen in \cref{fig:bpsuper}.

The coefficients $\sigma_k$ are the transseries parameters which encode the initial data. For the system \cref{eq:eoms}, one has four real integration constants. One of them survives in the perturbative series \cref{eq:asymptotic}, while the remaining ones are exponentially suppressed at late times. 
The exponential corrections reflect the quasinormal modes of the theory: one (corresponding to $k=0$ in \cref{eq:transseries-temp}) is a purely damped mode which is always present in MIS theory. The remaining two are not purely damped in general: they are complex conjugate and introduce oscillatory behavior of solutions whenever their real part does not vanish. 

\begin{figure}
    \centering
\includegraphics[width=0.7\linewidth]{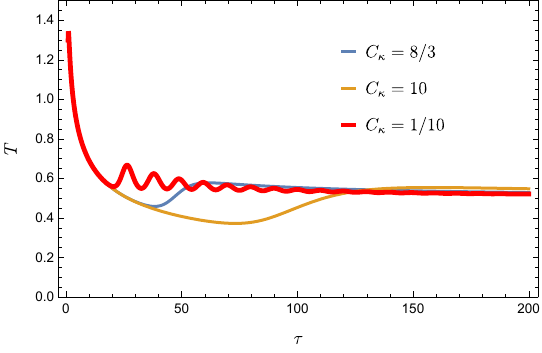}\\
 \vspace{1em}
\includegraphics[width=0.7\linewidth]{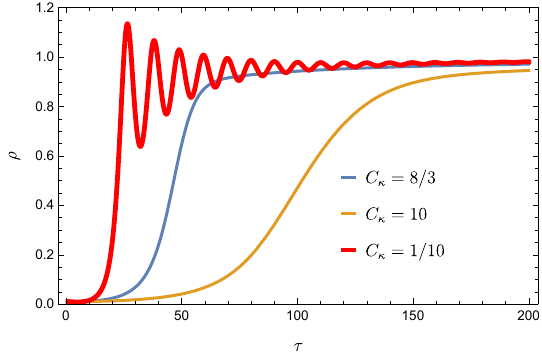}
    \caption{The temperature and the condensate evaluated by numerically solving \cref{eq:eoms} for the same set of initial conditions and parameters, 
    for three characteristic values of the condensate relaxation rate $C_\kappa$. Note that the critical value of $C_\kappa$ corresponds to $8/3$ with the choice of parameters \cref{eq:nice-parameters}.
    }
\label{fig:oscillations}
\end{figure}

\begin{figure}
    \centering
\includegraphics[width=0.7\linewidth]{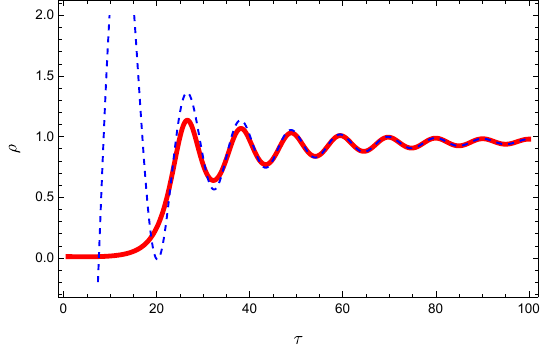}
    \caption{
   The red curve is 
    the numerical solution with $C_\kappa=1/10$ in \cref{fig:oscillations}, while the blue dashed curve is 
    the result of fitting $\Lambda$ and the transseries parameters $\sigma_k$ appearing in \cref{eq:transseries-temp}. 
 Note that the oscillations are due entirely to the exponential contributions to \cref{eq:transseries-temp} and is completely invisible to the perturbative series.
    }
    \label{fig:plofit}
\end{figure}

From \cref{eq:tspm} it follows that
there is a critical value of $C_\kappa$, namely
$C_\kappa^{\rm (crit)} = 8/3$ with the choice of parameters \cref{eq:nice-parameters}, beyond which the transseries contains no oscillating
contributions; a formula valid for arbitrary parameters is given in \Cref{app:general}.
This is a definite prediction coming out of the asymptotic
analysis. It can be verified by inspecting numerical solutions of \cref{eq:eoms} for various values of the condensate diffusion parameter $C_\kappa$, as illustrated in \cref{fig:oscillations}. In fact, the form of \cref{eq:transseries-temp} describes the late time behavior of the solutions rather well, as seen in \cref{fig:plofit}. 
To obtain this plot, the infinite series were truncated at order $1/\tau^3$ in the perturbative sector and at order $1/\tau$ in the first transseries sector. Note that the transseries parameters need to satisfy $\sigma_+^*=\sigma_-$ so as to render a real solution.

The presence of 
the oscillations discussed above
could leave an observable imprint in the spectra of hadrons emerging from a heavy ion collision. This possibility 
also highlights the significance of the condensate relaxation rate which plays a key role in determining whether this phenomenon occurs. It would be very interesting to understand the microscopic origin of this quantity.
We would also like to emphasize 
that the oscillations of the condensate are completely distinct from oscillations of the phase, which are expected to leave an imprint of the emitted pion spectra~\cite{Florio:2025lvu,Florio:2025zqv}. The condensate oscillations 
discussed here affect not just the condensate, but also the temperature
and the pressure anisotropy. Given that the latter couples directly to the perturbation of the post-freezeout distribution function~\cite{Teaney:2003kp}, one would expect that these oscillations would affect all particle spectra.

\section{Large order behavior}\label{sec:large}

We now turn to the infinite series $\Phi_k$ appearing in the transseries solution \cref{eq:transseries-temp}.
As in similar situations (see, e.g.~the reviews~\cite{Florkowski:2017olj,Soloviev:2021lhs,Jankowski:2023fdz}), these are formal asymptotic series and need to be resummed using appropriate techniques, or truncated. In the present case there is a novel feature in the form of the logarithmic terms seen in \cref{eq:transseries-temp} that require additional attention. In this article we will only give a brief account of these issues, leaving a more complete analysis to  future work. 

\begin{figure}
    \centering
 \includegraphics[width=0.7\linewidth]{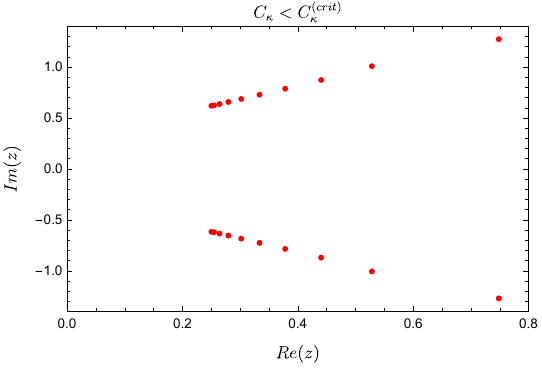}\\  
 \vspace{1em}
 \includegraphics[width=0.7\linewidth]{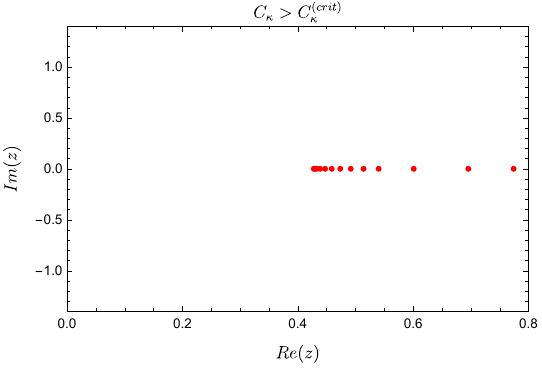} 
    \caption{
    The poles of the analytically continued Borel transform of the series $t_{k, 0}$ with $ k=1,\dots, 58$. The condensation of poles is indicative of branch point singularities. The pattern depends on whether the value of $C_\kappa$ exceeds the critical value defined in the text, see \cref{eq:crit-kappa}. 
    }  \label{fig:bpsuper}
\end{figure}

The formal series $\Phi^{(0)}$ 
can be grouped into log sectors as explained in  \Cref{sec:frozen} in the case of constant condensate. The series appearing in the lowest few log sectors can easily be calculated and turn out to diverge factorially. This is almost certainly true for all sectors. 
As an illustration, we now consider
the formal series in
the sector with $m=0$ in \cref{eq:transseries-temp} 
\begin{equation}
    S(\tau) \equiv \sum_{n=1}^\infty t_{n,0} \tau^{-n}.
\end{equation}
Its Borel transform is defined as 
\begin{equation}
    \tilde{S}(z) = \sum_{n=1}^\infty \frac{t_{n,0}}{n!}z^{n},
\end{equation}
where $z$ is a formal Borel plane variable. 
The Borel transform $\tilde{S}(z)$ 
converges near the origin and 
can be analytically continued away from it
using Pad\'e approximants to locate its singularities in the Borel plane. This analysis can be done in the same way as e.g.~in Refs.~\cite{Heller:2015dha,Aniceto:2015mto} and reveals dense sequences of poles indicative of branch points, seen in \cref{fig:bpsuper}. 
Analogous Borel-Pad\'e plots for $t_{k, 1}$ are almost indistinguishable from the plots for 
$t_{k, 1}$ 
and indicate branch points in the same locations. This is very important, since the cancellation of resummation ambiguities  relies on balancing contributions from different transseries sectors.

It is expected on general grounds that  the pattern of singularities seen in the Borel plane should match 
the exponential terms of the transseries~\cite{Heller:2015dha,Aniceto:2015mto,Aniceto:2018uik}. In \cref{fig:bpsuper} one can see the branch points at $z=\gamma_\pm$, but the branch point corresponding to $\gamma_0$ is not actually visible, since it is shadowed by the branch points closer to the origin of the Borel plane. This is a consequence of the specific values of the transport coefficients which were adopted for the calculation of the series. The branch point at $z=\gamma_0$ becomes visible in the Borel-Pad\'e plot in \cref{fig:bpsuper} if one increases the relaxation time $C_\tau$ so as to bring $\gamma_0$ closer  to the origin. In principle it should be visible for any value of $C_\tau$, but only with a much larger number of series coefficients. 

We note finally that the occurrence of oscillations (or a lack thereof) is reflected by the presence (or absence) of branch points away from the real axis, depending on the value of $C_\kappa$. For  $C_\kappa < C_\kappa^{\rm (crit)}$ there are two branch points (at conjugate positions in the Borel place), located at $\gamma_\pm$; for larger values of $C_\kappa$, these branch points approach the real axis and beyond the critical value 
all the branch points are on the real axis; in \cref{fig:bpsuper} only the one closest  to the origin is visible.

\section{Outlook}

We have studied the late time asymptotic behavior of Bjorken flow in a system describing an expanding QGP fluid interacting with a condensate resulting from the spontaneous symmetry breaking of a global $U(1)$ symmetry. We have found that the late time behavior of the effective temperature and pressure anisotropy are rather different from what is known from similar studies of MIS theory: the well-known Bjorken fall-off $T\sim\Lambda^{2/3}\tau^{-1/3}$ is replaced by a series involving integer powers and logs of the proper time. The series can be decomposed into log sectors, each containing a series with vanishing radius of convergence. 

The late time behavior of the hydrodynamic gradient expansion retains
information about the underlying microscopic theory which is not captured by
truncations of the perturbative series $\Phi^{(0)}(\tau)$. This is well known in the case of nonhydrodynamic
quasinormal modes of MIS theory~\cite{Heller:2015dha}, as well \symmm~\cite{Aniceto:2018uik} and various examples of kinetic
theory~\cite{Heller:2016rtz,Heller:2018qvh}. These modes play the role of a conduit by means of which information
about the initial state is relayed to the late time domain. In the case of the superfluid model discussed in this work, the nonhydrodynamic
modes include those associated with the spontaneous breaking of the $U(1)$ symmetry. 
Such contributions are exponentially suppressed relative to the perturbative sector, but they need not be small on physically relevant time scales. It can also happen that such terms carry qualitatively significant physical  effects. An example of this arises for instance in the context of QGP dynamics, where the dynamics in the transverse plane enters through exponentially suppressed corrections to Bjorken flow~\cite{An:2023yfq}.
This is also the case here: we have demonstrated that 
exponentially suppressed contributions that are beyond all orders of the gradient expansion can carry
qualitatively important physical effects, namely the oscillations associated
with relaxation of the condensate. 
Such oscillating quasinormal modes have previously been encountered in \symmm~\cite{Kovtun:2005ev,Aniceto:2015mto,Aniceto:2018uik,Spalinski:2018mqg}. 
Whether they occur in the superfluid system is determined by a
condition on the condensate relaxation rate, which is one of the main implications of
the form of the transseries established in the present study.

Our work leaves some open questions of mathematical nature. While we have established the form of the transseries and showed that it captures the late time behavior of numerical solutions, we have been working mostly with low-order truncations of the relevant asymptotic series. We have not made any attempts to sum these series or investigate resurgence relations which could be expected to relate expansion coefficients in different sectors (see e.g., Refs.~\cite{Aniceto:2015mto,Aniceto:2022dnm,Aniceto:2018bis}).  
It would be interesting to explore this in the future.

The present work was focused on the case of Bjorken flow, but it would also be interesting to investigate the asymptotic late time behavior in other situations. A natural conformal setting would be the de Sitter geometry first discussed in \cite{Gubser:2010ui,Gubser:2010ui}, corresponding to spherical slicing of dS$_3\times\mathbb{R}$, and recently generalized to include also hyperbolic slicing~\cite{Grozdanov:2025cfx} (see also recent applications \cite{Soloviev:2025uig,Martinez:2025jtv,Singh:2026wvf}).  Numerical studies of superfluid Gubser flow have already been  carried out in \cite{Buza:2024jxe}. We leave the asymptotic analysis of  flows in the Gubser or Grozdanov backgrounds to future work.

There are also interesting physical questions which can be addressed following the results reported here. We have considered the simplest possible example of QGP dynamics in the presence of a spontaneously broken symmetry. An important generalization would be to include the full content of the $SU(2)\times SU(2)$ chiral symmetry group of massless QCD, as well as 
explicit symmetry breaking in the potential. It would then be fascinating to study the possibility that damped oscillations of the condensate could be imprinted upon the spectrum of hadrons emerging from the collision. This would also include additional oscillations not studied here, due to the non-Abelian phase present in the $SU(2)\times SU(2)\sim O(4)$ case. Such oscillations have been seen in the pion correlation function in the instantaneously quenched scenario \cite{Florio:2025lvu,Florio:2025zqv}. This would provide strong motivation for exploring the microscopic physics responsible for the condensate relaxation rate, whose potential significance is perhaps the most important physical outcome of our work.

\acknowledgments{We would like to thank Lorenzo Gavassino  for helpful discussions.
We would also like to thank the European Centre for Theoretical Studies in Nuclear Physics and Related Areas (ECT*) in Trento (Italy) for their hospitality during the Workshop 
on ``Attractors and thermalization in nuclear collisions and cold atom systems''. 
AS is supported by funding from 
the project N1-0245 of Slovenian Research Agency (ARIS) and through the UL Startup project (UNLOCK) under contract no. SN-ZRD/22-27/510.
MS is supported by the National Science Centre, Poland, under Grant
No. 2021/41/B/ST2/02909.
}

\appendix
\crefalias{section}{appendix}

\section{Solutions} \label{app:solutions}
\allowdisplaybreaks
Explicitly, the asymptotic expansion of the temperature, condensate and  for the nice parameters in \cref{eq:nice-parameters} is given by
\begin{align}
   T(\tau)&=\frac{1}{2}+
   \frac{C_\eta \ln (\Lambda\tau )}{3 \tau }\\ \nonumber
   &+ \frac{1}{72\tau^{2}}\left[\left(-32 C_\eta^2+9 C_\eta C_\kappa+16
   C_\eta C_\tau\right)-\left(32 C_\eta^2+9 C_\eta C_\kappa\right) \ln (\Lambda \tau )-12 C_\eta^2 \ln^2(\Lambda \tau)\right]+\mathcal{O}(\tau^{-3}),\\
   \rho(\tau)&=1-\frac{C_\eta \ln (\Lambda\tau )}{3 \tau }\\
   &+\frac{1}{72\tau^2}\left(32 C_\eta^2+27 C_\eta C_\kappa-16
   C_\eta C_\tau
   +\left(32 C_\eta^2-27 C_\eta C_\kappa\right) \ln (\Lambda \tau )
   +8C_\eta^2 \ln^2(\Lambda \tau )\right)
   +\mathcal{O}(\tau^{-3}),\nonumber\\
   \chi(\tau)&=\frac{8 C_\eta}{3 \tau }-\frac{16 C_\eta}{9 \tau ^2}\left(C_\tau+C_\eta \ln (\Lambda \tau )\right)+\mathcal{O}(\tau^{-3}).
\end{align}

\section{Double logs}
\label{app:logs}

A series in logs and double logs appears in the context of 
the renormalization group equation in QCD~\cite{vanRitbergen:1997va,Gardi:1998qr,RevModPhys.77.837,Deur:2025rjo}, which we will briefly outline here. 
If we  consider the beta function in the usual RG approach, truncated to two loop order
\begin{align}\label{eq:beta-function}
    \mu g^\prime(\mu) =- \beta_0 g^2-\beta_1 g^3,
\end{align}
we can solve this asymptotically in $\tau\equiv \ln \mu/\Lambda$. 
We know that the (QCD) equation for the beta function should hold at low orders in perturbation theory the higher the energy is, which corresponds to the limit $\tau\rightarrow \infty$ and $g\rightarrow 0.$ Expanding in these limits, one finds an asymptotic solution of the form 
\begin{align}
     \tau g_0(\tau)&= \sum_{n=0}^\infty \sum_{m=0}^n a_{nm}\tau^{-n}\ln^m \tau,\\
     &=\frac{1}{\beta_0}
     +\frac{-2 \beta_1 \ln \beta_0
    -\beta_1 \ln \tau}{\beta_0^3 \tau}
    +\frac{\beta_1^2 \ln^2\tau
    +\left(4 \beta_1^2 \ln \beta_0 -\beta_1^2 \right) \ln \tau
    +\left(4 \ln^2\beta_0-2 \ln \beta_0-1\right)\beta_1^2}{\beta_0^5 \tau^2}+\ldots\nonumber
\end{align}
Curiously, in this case one can compute an exact closed form expression to the truncated equation \cref{eq:beta-function} in terms of the Lambert W function \cite{Gardi:1998qr}
\begin{align}
    g(\tau)&= -\frac{\beta_0/\beta_1}{1+W_0(\beta_1^{-1} e^{-1-\beta_1^{-1}\beta_0^2\tau})}.
\end{align}
Expanding order by order in large $\tau$, we find exact agreement with the asymptotic expansion. 

Another instance where a series of similar form has appeared is in recent studies of nonthermal attractors~\cite{Heller:2023mah}. There it arises as an asymptotic late time solution of an ordinary differential equation  which controls prescaling in a system undergoing homogeneous isotropization:
\begin{align}
   \frac{\beta
   ''(t)}{\beta (t)}= \frac{\beta '(t)^2}{\beta
   (t)^2}+\frac{14 \beta '(t)}{t}-\frac{(7 \beta (t)+1)^2}{t^2}.
\end{align}
By inspection, the leading asymptotic contribution is $\beta_\infty = -1/7$. Subsequent subleading terms follow
a double series in $\ln (Q t)$ and $\ln\ln (Q t)$, where $Q$ is an integration
constant. Upon setting $\tau = \ln(Q t)$, this double series takes the form of \cref{eq:toyasym}.

\section{General expressions}\label{app:general}

In the main text, we reported on the form of the transseries \cref{eq:transseries-temp} with a particular choice of parameters \cref{eq:nice-parameters}. For completeness, here we provide the transseries for $\rho$ 
with arbitrary parameters. The purely decaying mode is given by
\begin{align}
    \delta \rho &= \tau^{\beta}e^{-\frac{(m_0 T_c-\rho^2_0\lambda)}{2C_\tau}\tau -\frac{ \rho_0\lambda C_\eta}{3 m_0 C_\tau}\ln^2(\Lambda\tau)}(1-\frac{C_\eta^2 \lambda}{9 m_0 C_\tau}\frac{\ln^2(\Lambda\tau)}{\tau}+\ldots ),\\
    \nonumber
    \beta&=\frac{A}{B},\\
    A&=2 \lambda  \rho_0^2 m_0 ^4 \left(-120 T_c^4 (2 C_{\kappa}
   C_\tau-1)+3 \rho_0^2 m_0  T_c (5 C_{\kappa}
   C_\tau+1)
   -5 C_\tau^2 \rho_0^2 m_0 ^2+48 C_\tau^2
   m_0  T_c^3\right)\nonumber\\
   &+m_0 ^5 T_c \left(48 T_c^4 (2 C_{\kappa}
   C_\tau-1)-3 \rho_0^2 m_0  T_c (5 C_{\kappa}
   C_\tau+1)+4 C_\tau^2 \rho_0^2 m_0 ^2-48 C_\tau^2
   m_0  T_c^3\right)\nonumber\\
   &+48 \lambda ^4 \rho_0^8 m_0  \left(5 T_c (2
   C_{\kappa} C_\tau-1)+C_\tau^2 m_0 \right)-96 \lambda ^3
   \rho_0^6 m_0 ^2 T_c \left(5 T_c (2 C_{\kappa} C_\tau+C_\tau^2 m_0 \right)
   \nonumber\\
   &+48 \lambda^5 \rho_0^{10} (1-2
   C_{\kappa} C_\tau)-3 \lambda ^2 \rho_0^4 m_0 ^3
   \left(\rho_0^2 m_0  (5 C_{\kappa} C_\tau+1)-160 T_c^3
   (2 C_{\kappa} C_\tau-1)\right) \nonumber\\
   B&=6 \left(\lambda  \rho_0^2-m_0  T_c\right) (-m_0 ^4
   \left(T_c^4 (4-4 C_{\kappa} C_\tau)+C_\tau^2 \rho_0^2 m_0 ^2\right)
   -8 \lambda  \rho_0^2 m_0 ^3 T_c^2 \left(2
   T_c (C_{\kappa} C_\tau-1)+C_\tau^2 m_0 \right)\nonumber\\
   &-8
   \lambda ^3 \rho_0^6 m_0  \left(2 T_c (C_{\kappa} C_\tau-1)+C_\tau^2 m_0 \right)+8 \lambda ^2 \rho_0^4 m_0 ^2
   T_c \left(3 T_c (C_{\kappa} C_\tau-1)+2 C_\tau^2
   m_0 \right)+4 \lambda ^4 \rho_0^8 (C_{\kappa} C_\tau-1))\nonumber
\end{align}
Surprising as it may be, for the nice parameters \cref{eq:nice-parameters}, we find that the above expression reduces to $\beta=-7/3$ for arbitrary transport coefficients.

The oscillatory modes are
\begin{align}
    \delta \rho &= \tau^{\beta}e^{-\gamma_\pm\tau -\gamma^\prime_\pm\ln^2(\Lambda\tau)}(1+
    \ldots ),\\
    \gamma_\pm &= -\frac{C_\kappa \left(\lambda  \rho_0^2-\sigma 
T_c\right)^3\pm \sqrt{C_\kappa^2 \left(\lambda  \rho_0^2-\sigma 
   T_c\right)^6-\rho_0^2 \sigma ^2 \left(\lambda  \rho_0^2-\sigma
    T_c\right)^2 \left(8 \lambda ^3 \rho_0^4+\sigma ^4+8 \lambda  \sigma
   ^2 T_c^2-16 \lambda ^2 \rho_0^2 \sigma 
   T_c\right)}}{2 \sigma  \left(\lambda  \rho_0^2-\sigma 
   T_c\right)^2},\nonumber\\
   \gamma^\prime_\pm &=\frac{C_\eta \gamma_\pm E}{6 \rho_0 \left(\lambda  \rho_0^2-\sigma  T_c\right) F},\\
   E&= 
   -4 \gamma_\pm^2 \sigma 
   (C_\kappa C_\tau+1) \left(\sigma  T_c-\lambda 
   \rho_0^2\right)^3 \left(5 \sigma  T_c-13 \lambda 
   \rho_0^2\right)-20 C_\kappa\gamma_\pm \left(3
   \lambda  \rho_0^2-\sigma  T_c\right) \left(\lambda 
   \rho_0^2-\sigma  T_c\right)^4\nonumber\\
   &+4 C_\tau
   \gamma_\pm^3 \sigma ^2 \left(5 \sigma  T_c-11 \lambda  \rho_0^2\right) \left(\lambda  \rho_0^2-\sigma 
   T_c\right)^2-C_\tau \gamma_\pm \rho_0^2 \sigma
   ^2 \Big(104 \lambda ^4 \rho_0^6+\lambda  \sigma ^3 \left(9
   \rho_0^2 \sigma -56 T_c^3\right)
   \nonumber\\
   &+216 \lambda ^2 \rho_0^2 \sigma ^2 T_c^2-264 \lambda ^3 \rho_0^4 \sigma 
   T_c-7 \sigma ^5 T_c\Big)-\rho_0^2 \sigma  \Big(120
   \lambda ^4 \rho_0^6+\lambda  \sigma ^3 \left(11 \rho_0^2
   \sigma -56 T_c^3\right)+232 \lambda ^2 \rho_0^2 \sigma ^2
   T_c^2\nonumber\\
   &-296 \lambda ^3 \rho_0^4 \sigma  T_c-7 \sigma ^5
   T_c\Big) \left(\lambda  \rho_0^2-\sigma 
   T_c\right)\nonumber\\
    F&=12 \gamma_\pm^2 \sigma 
   (C_\kappa C_\tau+1) \left(\lambda  \rho_0^2-\sigma  T_c\right)^3+8 C_\kappa\gamma_\pm
   \left(\lambda  \rho_0^2-\sigma  T_c\right)^4
   +16
   C_\tau \gamma_\pm^3 \sigma ^2 \left(\lambda  \rho_0^2-\sigma  T_c\right)^2\nonumber\\
   &+2 C_\tau \gamma_\pm
   \rho_0^2 \sigma ^2 \left(8 \lambda ^3 \rho_0^4+\sigma
   ^4+8 \lambda  \sigma ^2 T_c^2-16 \lambda ^2 \rho_0^2 \sigma
    T_c\right)+\rho_0^2 \sigma  \left(8 \lambda ^3 \rho_0^4+\sigma ^4+8 \lambda  \sigma ^2 T_c^2-16 \lambda ^2 \rho_0^2 \sigma  T_c\right) \left(\lambda  \rho_0^2-\sigma 
   T_c\right),
   \nonumber\\
   \beta&=\frac{C}{6D},\\
   C&=4 \lambda  \rho_0^2 \sigma ^4 \left(2 \gamma_\pm T_c^2
   \left(4 \gamma_\pm T_c (11 C_\kappa C_\tau+8)+5 C_\kappa T_c^2-42 C_\tau
   \gamma_\pm^2\right)+3 C_\kappa\gamma_\pm \lambda 
   \rho_0^4+4 \lambda  \rho_0^2 T_c^2 (50
   T_c-33 C_\tau \gamma_\pm)\right)\nonumber\\
   &-8 \lambda ^4
   \rho_0^8 \sigma  \left(\gamma_\pm^2 (11 C_\kappa
   C_\tau+8)+5 C_\kappa\gamma_\pm T_c+16 \lambda
    \rho_0^2\right)\nonumber\\
    &+16 \lambda ^3 \rho_0^6 \sigma ^2
   \left(C_\tau \gamma_\pm \left(22 C_\kappa
   \gamma_\pm T_c-7 \gamma_\pm^2-11 \lambda  \rho_0^2\right)+T_c \left(5 C_\kappa\gamma_\pm
   T_c+16 \gamma_\pm^2+35 \lambda  \rho_0^2\right)\right)\nonumber\\
    &-16
   \lambda ^2 \rho_0^4 \sigma ^3 T_c \left(T_c \left(5
   C_\kappa\gamma_\pm T_c+24 \gamma_\pm^2+60 \lambda 
   \rho_0^2\right)-3 C_\tau \gamma_\pm \left(-11
   C_\kappa\gamma_\pm T_c+7 \gamma_\pm^2+11 \lambda 
   \rho_0^2\right)\right)\nonumber\\
    &+\sigma ^6 \left(3 \gamma_\pm \rho_0^2 (C_\tau \gamma_\pm (C_\kappa
   T_c-\gamma_\pm)+2 T_c (2 C_\kappa
   T_c+\gamma_\pm))+2 \lambda  \rho_0^4 (30 T_c-13
   C_\tau \gamma_\pm)+48 T_c^5\right)
   \nonumber\\
    &+\sigma ^5 \Big(-8
   \gamma_\pm T_c^3 (C_\tau \gamma_\pm (11 C_\kappa T_c-14 \gamma_\pm)+T_c (C_\kappa T_c+8
   \gamma_\pm))\nonumber\\
    &-3 \gamma_\pm \lambda  \rho_0^4 (\gamma_\pm
   (C_\kappa C_\tau+2)+8 C_\kappa T_c)+16
   \lambda  \rho_0^2 T_c^3 (11 C_\tau
   \gamma_\pm-20 T_c)-36 \lambda ^2 \rho_0^6\Big)\nonumber\\
    &+8
   C_\kappa\gamma_\pm \lambda ^5 \rho_0^{10}+4
   \rho_0^2 \sigma ^7 T_c (5 C_\tau \gamma_\pm-6
   T_c),\nonumber\\
   D&= 4 C_\kappa\gamma_\pm \left(\lambda 
   \rho_0^2-\sigma  T_c\right)^4 \left(3 C_\tau
   \gamma_\pm \sigma +2 \lambda  \rho_0^2-2 \sigma 
   T_c\right)\nonumber\\
    &+\sigma  \left(\lambda  \rho_0^2-\sigma 
   T_c\right) \Big(16 C_\tau \gamma_\pm^3 \sigma 
   \left(\lambda  \rho_0^2-\sigma  T_c\right)^2+2 C_\tau \gamma_\pm \rho_0^2 \sigma  \left(8 \lambda ^3 \rho_0^4+\sigma ^4+8 \lambda  \sigma ^2 T_c^2-16 \lambda ^2 \rho_0^2 \sigma  T_c\right)\nonumber\\
    &+12 \gamma_\pm^2 \left(\lambda 
   \rho_0^2-\sigma  T_c\right)^3+\rho_0^2
   \left(\lambda  \rho_0^2-\sigma  T_c\right) \left(8 \lambda
   ^3 \rho_0^4+\sigma ^4+8 \lambda  \sigma ^2 T_c^2-16 \lambda
   ^2 \rho_0^2 \sigma  T_c\right)\Big).\nonumber
\end{align}
Finally, for general values of the potential parameters, the
critical value of the transport coefficient is
\begin{align}\label{eq:crit-kappa}
 (C_\kappa^{\rm (crit)})^2&=\frac{\rho_0^2 m_0^2  \left(m_0^4+8 \lambda  \left(\lambda  \rho_0^2-m_0  T_c\right)^2\right)}{\left(\lambda  \rho_0^2-m_0 
   T_c\right)^4}
\end{align}
This reduces to $8/3$ for the nice parameters in \cref{eq:nice-parameters}.

\bibliography{main}
\bibliographystyle{jhep}

\appendix

\end{document}